\documentclass[showpacs,preprintnumbers,amsmath,amssymb]{revtex4}

\usepackage{dcolumn}
\usepackage{bm}

\newcommand{\msbar}{\overline{\mbox{MS}}}

\newcommand{\beq}{\begin{equation}}
\newcommand{\eeq}{\end{equation}}
\newcommand{\bea}{\begin{eqnarray}}
\newcommand{\eea}{\end{eqnarray}}

\newcommand{\BreakI}{ \right. \nonumber \\ &{}& \left. }
\newcommand{\BreakII}{ \right. \right.  \nonumber \\ &{}& \left.\left. }

\newcommand{\lsb}{\left[}
\newcommand{\rsb}{\right]}

\begin{document}
\hfill THEP        02/17\\
\mbox{}
\hfill   TTP02--41\\
\mbox{}
\hfill  hep-ph/0212299 \\

\title{Towards Order $\alpha_s^4$ Accuracy in $\tau$-decays\\}

\author{P.A.~Baikov}
 \email{baikov@theory.sinp.msu.ru}
\affiliation{
Institute of Nuclear Physics, 
Moscow State University
Moscow~119992, Russia\\
}%

\author{K.G.~Chetyrkin}
 \altaffiliation[On leave from ]
{Institute for Nuclear Research of the Russian Academy of Sciences, 
Moscow, 117312, Russia} 
\email{chet@particle.physik.uni-karlsruhe.de}
\affiliation{%
Fakult{\"a}t f{\"u}r Physik
 Albert-Ludwigs-Universit{\"a}t Freiburg,
D-79104 Freiburg, Germany\\
}

\author{J.H.~K{\"u}hn}
 \email{Johann.Kuehn@physik.uni-karlsruhe.de}
\affiliation{Institut f\"ur Theoretische Teilchenphysik,
    Universit\"at Karlsruhe,
     D-76128 Karlsruhe, Germany\\
}%

\date{\today}

\begin{abstract}
Recently computed terms of orders $\mathcal{O}(\alpha_s^4 n_f^2)$ in
the perturbative series for the $\tau$ decay rate, and similar (new)
strange quark mass corrections, are used to discuss the validity of
various optimization schemes. The results are then  employed  to arrive at
improved predictions for the complete terms order
$\mathcal{O}(\alpha_s^4)$ and $\mathcal{O}(\alpha_s^5)$ in the
massless limit as well as for  terms due to the strange quark mass.
Phenomenological implications are presented.
\end{abstract}

\maketitle

\section{Introduction}\label{sec:intro}
The dependence of the $\tau$-decay rate on the strong coupling
$\alpha_s$ has been used for a determination of $ \alpha_s $ at lower
energies, with the most recent results of $0.334 \pm 0.007_\mathrm{exp} \pm
0.021_\mathrm{theo}$ and  $0.3478\pm 0.009_\mathrm{exp} \pm 0.019_\mathrm{theo}$ 
by the ALEPH
\cite{exp1} and OPAL \cite{exp2} collaborations. After evolution up
to higher energies these results agree remarkably well with determinations based on
the hadronic $Z$ decay rate. In view of the relatively large value of
$\alpha_s(M_{\tau}) $ estimates for the yet unknown terms of higher
orders play an important role in current determinations of $ \alpha_s $ at
low energies. This is in contrast to high energy measurements, where
the uncertainty from terms of order $ \alpha_s^4 $ of 0.001 to 0.002
\cite{phys_report} is somewhat less or at most comparable  
 to the present experimental errors.

The situation is even more problematic for the determination of the
strange quark mass from the Cabbibo suppressed $\tau$
decays. Perturbative QCD corrections affecting the $m_s^2$ term are
extremely large and contributions from increasing powers of $ \alpha_s $ are
barely decreasing, which casts doubts on our ability to extract a
reliable result for $m_s$ from this (in principle) clean and
straightforward measurement 
\cite{Maltman:1998qz,Chetyrkin93,Chetyrkin93-update,Chetyrkin:1998ej,pichprades,%
krajms,chenpich,Maltman:2002wb}.

Partial results of order $\alpha_s^4$ for the absorptive part of the
massless vector and scalar correlators have been obtained recently
\cite{Baikov:2001aa}, namely terms proportional to $n_f^2$, where
$n_f$ denotes the number of massless fermion species. These allow to test
two popular optimization schemes --- based on the principles of
``minimal sensitivity'' (PMS) and of fastest apparent convergence
(FAC) \cite{ste,gru,krasn} 
--- which have been used to predict yet
uncalulated higher order terms \cite{KS,Kataev:1994wd}.

It will be demonstrated that the predictions of both schemes
(coinciding  at $ \alpha_s ^4$) for the coefficient of order
$n_f^2\alpha_s^4$ are in reasonable agreement with our calculations,
which are then used to predict the complete fixed order (FO) and the
"contour improved" (CI) \cite{tau:resum,DP} $\mathcal{O}(\alpha_s^4)$
contributions to the $\tau$ decay rate. Employing the four-loop QCD beta
function in combination with improved $ \alpha_s ^4$ terms a rough estimate
even for $\mathcal{O}(\alpha_s^5)$ terms can be obtained.  The results
lead to fairly stable values for $\alpha_s$ consistent with current
analysises.

Unfortunately, no essential decrease of the difference between the
central values $ \alpha_s (M_{\tau})$ as obtained with FI and CI approaches
is observed after including order $ \alpha_s ^4$ and $ \alpha_s ^5$ corrections,
assuming for the moment that these estimates are indeed correct.  On
the other hand, the theoretical uncertainty assigned to
$ \alpha_s (M_{\tau})$ {\em within} each method according to standard
techniques  does decrease significantly.

The implication of this approach for the extraction of $m_s$ from
Cabbibo suppressed decays will be investigated along the same lines.
New results will be presented for the terms of order $ n_f  \alpha_s ^3\,
m_s^2$ in the total rate.  In this case the agreement between FAC/PMS
predictions and our results is quite encouraging and naturally
suggests the use of the former as a reliable prediction for the
complete $  \alpha_s ^3 m_s^2$ term. Following an approach discussed in
\cite{Kataev:1994wd}, even a rough estimate $m_s^2\alpha_s^4$ 
terms can be  obtained from these considerations.

However, the rapid increase of the coefficients indicates that
the inherent uncertainty of the present $m_s$ determinations will 
not necessarily decrease with  inclusion
of the higher orders. As we will see, the situation is
somewhat better for the spin one contribution if considered
separately.

\section{Generalities}

We start with the well-known representation 
\cite{SchTra84,Bra88,NarPic88,Bra89,BraNarPic92} 
of the
tau-lepton hadronic rate
as the contour integral along a
circle C of  radius
$|s|=M_{\tau}^2$
 \begin{equation}   
R_{\tau}=6i\pi\int_{|s|=M_{\tau}^2}\frac{ds}{M_{\tau}^2}
\left(1-\frac{s}{M_{\tau}^2}\right)^2
\left[\Pi^{[2]}(s)
-\frac{2}{M_{\tau}^2}\Pi^{[1]}(s)\right]
\label{pi1andpi2}
{}.
 \end{equation} 

Here $\Pi^{[1]}$ and $\Pi^{[2]}$
are proper flavour combinations of the polarization   operators
appearing in the  decomposition of the  correlators 
of  vector and axial vector currents of light quarks
\begin{equation}
\begin{array}{ll}
\Pi^{V/A}_{\mu\nu,ij}(q,m_i,m_j,m{},\mu,\alpha_s) & =
\displaystyle i \int dx e^{iqx}
\langle
T[\, j^{V/A}_{\mu,ij}(x) (j^{V/A}_{\nu,ij})^{\dagger} (0)\, ] \rangle
\\ &  = \displaystyle 
g_{\mu\nu}  \Pi^{[1]}_{ij,V/A}(q^2) 
      +  q_{\mu}q_{\nu} 
  \Pi^{[2]}_{ij,V/A}(q^2)
{}  
\end{array} 
\label{correlator}
\end{equation}
with  $m^2 = \sum_{f=u,d,s}  m_f^2$
and  $j^{V/A}_{\mu,ij} = \bar{q}_i\gamma_{\mu}(\gamma_5) q_j$.
The  two (generically different) quarks with
masses $m_i$ and  $m_j$ are denoted by  $q_i$ and $q_j$  respectively. 

For the  case of the $\tau$-lepton  the relevant combinations
of quark flavours are $ij  = ud$ and   $ij  = us$. 
The polarization functions  $\Pi^{[l]}_{V/A}, \ \ l=1,2$                               
are  conveniently represented  in                                        
the form  ($Q^2  = -q^2$)
\begin{equation}                                                            
(Q^2)^{(l-2)}\Pi^{[l]}_{us,V/A}(q^2)                                   
=                                                                          
\frac{3}{16\pi^2}\Pi^{[l]}_{V/A,0}(\frac{\mu^2}{Q^2}, \alpha_s)    
+                                                      
\frac{3}{16 \pi^2}
\sum_{D \ge 2} Q^{-D}\Pi^{[l]}_{V/A,D}(\frac{\mu^2}{Q^2},
m_s^2,\alpha_s)      
{}.                                                                         
\label{mass-exp}                                                            
\end{equation}                                                              
Here the first term on the rhs  corresponds  to the massless               
limit while the first term in the sum 
stands for quadratic mass corrections. We neglect the
masses of $u$ and $d$ quarks.   Therefore in  perturbative QCD 
$\Pi^{[l]}_{V} = \Pi^{[l]}_{A}$
and we  will  often omit the subscript $V/A$ in  the following.
Current conservation implies:
$ 
\Pi^{[1]}_{0}  = \Pi^{[2]}_{0}
$.

The full tau-lepton hadron rate  $R_{\tau}$ can be presented as a sum of 
spin 1 and spin 0 parts, viz.
 \begin{equation}  
\label{spin:decomp}
 \begin{array} {ll}
 \displaystyle 
R^{(1)}_{\tau} &= 
 \displaystyle 
6i\pi\int_{|s|=M_{\tau}^2}\frac{ds}{M_{\tau}^2}
\left(1-\frac{s}{M_{\tau}^2}\right)^2
\left[ \left(1+2\frac{s}{M_{\tau}^2}\right) \Pi^{(1)}(s)
+\Pi^{[1]}(0)/s
\right]
{},
\\
 \displaystyle 
R^{(0)}_{\tau} &= 
 \displaystyle 
6i\pi\int_{|s|=M_{\tau}^2}\frac{ds}{M_{\tau}^2}
\left(1-\frac{s}{M_{\tau}^2}\right)^2
\left[\Pi^{(0)}(s)
-\Pi^{[1]}(0)/s
\right]
{}.
 \end{array} 
 \end{equation} 
where 
 \begin{equation} 
\Pi^{(1)}  = -\Pi^{[1]}/q^2, \ \  
\Pi^{(0)}  = \Pi^{[2]} +    \Pi^{[1]}/q^2
{}.
\label{Pi12TOPi01}
 \end{equation} 
and the contribution of the singularity at the origin 
(proportional to $\Pi^{[1]}(0)$) has to be included.
A nonvanishing  value of $\Pi^{[1]}(0)$   
is   a nonperturbative 
constant. 

On the other hand, the unknown constant drops out if one
considers moments 
 \begin{equation} 
R^{(1,0)k,l}_{\tau}(s_0) =  
\int_0^{s_0}
\frac{ds}{M_{\tau}^2}
\left(1-\frac{s}{M_{\tau}^2}\right)^{k} 
\left(\frac{s}{M_\tau^2}\right)^l
\frac{d R^{(1,0)}_\tau }{ds}
{},
\label{def:moments}
 \end{equation} 
with $k \ge 0, \ \ l \ge 1$. 
(Note that  the moments introduced in \cite{DP} 
are related to  ours as
$
R^{kl}_\tau = R^{(1)k,l}_\tau +  R^{(0)k,l}_\tau $.)

The decay rate 
$R_{\tau}$ may be expressed as the sum of different contributions
corresponding to Cabibbo suppressed or allowed decay modes, vector or
axial vector contributions and the mass dimension of the corrections
 \begin{equation}   \label {7}                                                              
R_{\tau} = R_{\tau,V} + R_{\tau,A} + R_{\tau,S}                           
 \end{equation}                                                                       
with                                                                      
 \begin{equation}   \label {8}                                                              
 \begin{array} {ll}  \displaystyle                                                                
R_{V} =                                                                   
&  \displaystyle                                                                      
\frac{3}{2}|V_{ud}|^2 \left( 1 + \delta_{0} + \sum_{D=2,4,\dots}        
\delta_{V,ud,{D}} \right),  \\                                          
R_{A} =                                                                   
&  \displaystyle                                                                      
\frac{3}{2}|V_{ud}|^2 \left( 1 + \delta_{0} + \sum_{D=2,4,\dots}        
\delta_{A,ud,{D}} \right),  \\                                          
R_{S} =                                                                   
&  \displaystyle                                                                      
3|V_{us}|^2 \left( 1 + \delta_{0} + \sum_{D=2,4,\dots}                  
\delta_{us,{D}} \right).                                                
 \end{array}                                                                        
 \end{equation}                                                                       
Here $D$ indicates the mass dimension of the fractional corrections
$\delta_{V/A,ij,{D},}$ and $\delta_{ij,{D}}$ denotes the average of
the vector and the axial vector contributions:
$\delta_{ij,{D}}=(\delta_{V,ij,{D}}+\delta_{A,ij,{D}})/2$.  If 
a decomposition into different spin/parity contributions is made or a
particular pattern of  moments is considered then we will use the
corresponding obvious generalization of (\ref{8}). 
For instance, 
 \begin{equation} 
R^{(1)kl}_{S,V} = a_{kl} \, |V_{us}|^2 \left( 1 + \delta^{kl}_0 + \sum_{D=2,4,\dots}                  
\delta^{(1),kl}_{V,us,D} \right)
{},                                                
 \end{equation}    
and 
 \begin{equation} 
R^{(0)kl}_{S,V} =                                                                   
|V_{us}|^2 \left( \sum_{D=2,4,\dots}                  
\delta^{(0),kl}_{V,us,D} \right).                                                
 \end{equation} 
Thus, in our notation we have the relation
 \begin{equation} 
\delta^{kl}_{V,us,{2}}
= a_{kl} \delta^{(1),kl}_{V,us,2} + \delta^{(0),kl}_{V,us,2}
\label{}
{}.
 \end{equation}

The integral in 
eq.~(\ref{pi1andpi2}) 
is, obviously, insensitive,
to  the $Q^2$-independent terms in the polarization functions
$ \Pi^{[1]}_{0}$ and $\Pi^{[1]}_{2} {}$.
This means that without  loss of generality we may 
deal with the corresponding (Adler) $D$-functions, viz.
\[
D^{[1]}_0(Q^2) \equiv -\frac{3}{4}Q^2 \frac{ {\rm d}}{{\rm d} Q^2}
\Pi^{[1]}_{0} , 
\   \  \
D^{[1]}_2(Q^2) \equiv -\frac{1}{2}Q^2 \frac{ {\rm d}}{{\rm d} Q^2}
\Pi^{[1]}_{2} 
{}.
\]
An important property of the functions  $ D^{[1]}_0, D^{[1]}_2$ and
$ \Pi^{[2]}_{2} $ is their scale independence,  which  implies that 
they are directly related to measurements.

The Adler functions $ D^{[1]}_0$ and $D^{[1]}_2$ have been
calculated with $\mathcal{O}(\alpha_s^3)$ accuracy, the polarization
function $\Pi^{[2]}_{2}$, however, to $\mathcal{O}(\alpha_s^2)$ only
(see  \cite{Chetyrkin:1998ej} and references therein).

The (apparent) convergence of the perturbative series for $ D^{[1]}_0$
is  acceptable, the one for $D^{[1]}_2$ and $ \Pi^{[2]}_{2} $
is at best marginal. This has led to significant theoretical
uncertainties in extracting $\alpha_s$ and to a fairly unstable
behaviour in   extraction of $m_s$ from
$\tau$-decays.

To improve the situation, we have computed the two leading terms in the
large $n_f$ expansion of the next order, i.e. terms of order
$\mathcal{O}(n_f^3\alpha_s^4)$, $\mathcal{O}(n_f^2\alpha_s^4)$ to $
D^{[1]}_0$, $ D^{[1]}_2$ and $\mathcal{O}(n_f^2\alpha_s^3)$,
$\mathcal{O}(n_f\alpha_s^3)$ to $\Pi^{[2]}_2$.  Our results are
described in the following section.

\section{Fixed order results in $ \alpha_s ^3$ and  $ \alpha_s ^4$}

Using the technique described in \cite{bai1,bai2,Baikov:2001aa} 
and the parallel version of FORM \cite{Fliegner:2000uy,Vermaseren},
the leading and subleading (in $n_f$) terms of the next  order in the
perturbative series for $ D^{[1]}_0$, $ D^{[1]}_2$ and $\Pi^{[2]}_2$
have  been obtained in the standard $\msbar$  renormalization scheme \cite{ms,msbar}:
\bea 
 D^{[1]}_{0} &=& 1 +\, a_s + a_s^2 \left\{
 \lsb -\frac{11}{12}+\frac{2}{3} \,\zeta_{3} \rsb  \,n_f +\frac{365}{24}-11 \,\zeta_{3}\right\}
\nonumber
 \\ &+& a_s^3 \left\{
 \lsb \frac{151}{162}-\frac{19}{27} \,\zeta_{3} \rsb  \, n_f^2 +\lsb -\frac{7847}{216}+\frac{262}{9} \,\zeta_{3}-\frac{25}{9} \,\zeta_{5} \rsb  \,n_f \BreakI 
\hspace{1cm}
+\frac{87029}{288}-\frac{1103}{4} \,\zeta_{3}+\frac{275}{6} \,\zeta_{5}\right\}\nonumber
   \\ &+&a_s^4\left\{
 \lsb -\frac{6131}{5832}+\frac{203}{324} \,\zeta_{3}+\frac{5}{18} \,\zeta_{5} \rsb  \, n_f^3 +\lsb \frac{1045381}{15552}+\frac{5}{6} \,\zeta_3^2\BreakII 
\hspace{1cm}
-\frac{40655}{864} \,\zeta_{3}-\frac{260}{27} \,\zeta_{5} \rsb  \, n_f^2 + d^{[1]4}_{0,1} \,n_f +d^{[1]4}_{0,0}\right\} \nonumber
   \\ &=&   1 +\, a_s + a_s^2 \left\{
-0.1153 \,n_f +1.986\right\}
\nonumber
 \\ &+& a_s^3 \left\{
0.08621 \, n_f^2-4.216 \,n_f +18.24\right\}\nonumber
   \\ &+&a_s^4\left\{
-0.01009 \, n_f^3 + 1.875 \, n_f^2 + d^{[1]4}_{0,1} \,n_f +d^{[1]4}_{0,0}\right\} {},  
\label{D1m0:exact} \end{eqnarray}
\bea 
 D^{[1]}_{2} =   m_s^2 \left(\rule{0mm}{7mm} \right. 1 &+& \frac{5}{3}\, a_s + a_s^2 \left\{
 \lsb -\frac{11}{8}+\frac{2}{3} \,\zeta_{3} \rsb  \,n_f +\frac{5185}{144}-\frac{39}{2} \,\zeta_{3}\right\}
\nonumber
 \\ &+& a_s^3 \left\{
 \lsb \frac{8671}{11664}-\frac{13}{27} \,\zeta_{3} \rsb  \, n_f^2 +\lsb -\frac{44273}{972}+\frac{3257}{81} \,\zeta_{3}\BreakII 
\hspace{1cm}
-\frac{5}{6} \,\zeta_{4}-\frac{1265}{81} \,\zeta_{5} \rsb  \,n_f +\frac{2641517}{5184}-\frac{131275}{216} \,\zeta_{3}+\frac{12845}{36} \,\zeta_{5}\right\}\nonumber
   \\ &+&a_s^4\left\{
 \lsb -\frac{396781}{559872}+\frac{461}{1296} \,\zeta_{3}-\frac{1}{48} \,\zeta_{4}+\frac{5}{18} \,\zeta_{5} \rsb  \, n_f^3 +\lsb \frac{61913567}{1119744}-\frac{59}{54} \,\zeta_3^2\BreakII 
\hspace{1cm}
-\frac{352549}{7776} \,\zeta_{3}+\frac{67}{96} \,\zeta_{4}+\frac{22859}{3888} \,\zeta_{5} \rsb  \, n_f^2 + d^{[1]4}_{2,1} \,n_f +d^{[1]4}_{2,0}\right\} \left.\rule{0mm}{7mm} \right)
 \nonumber
   \\  \phantom{\Pi^{[2]}_{2}} =  m_s^2 \left(\rule{0mm}{7mm} \right.1 &+&1.667\, a_s + a_s^2 \left\{
-0.5736 \,n_f +12.57\right\}
\nonumber
 \\ &+& a_s^3 \left\{
0.1646 \, n_f^2-14.31 \,n_f +149.\right\}\nonumber
   \\ &+&a_s^4\left\{
-0.01563 \, n_f^3 + 6.067 \, n_f^2 + d^{[1]4}_{2,1} \,n_f +d^{[1]4}_{2,0}\right\} \left.\rule{0mm}{7mm} \right)  {},  
\label{D1m2:exact} \end{eqnarray}
\bea 
 \Pi^{[2]}_{2} = -4 m_s^2 \left(\rule{0mm}{7mm} \right. 1 &+&\frac{7}{3}\, a_s + a_s^2 \left\{
 \lsb -\frac{25}{24}-\frac{2}{9} \,\zeta_{3} \rsb  \,n_f +\frac{15331}{432}+\frac{359}{54} \,\zeta_{3}-\frac{520}{27} \,\zeta_{5}\right\}
\nonumber
 \\ &+& a_s^3 \left\{
 \lsb \frac{2131}{11664}+\frac{19}{81} \,\zeta_{3} \rsb  \, n_f^2 +\lsb -\frac{68135}{1944}-\frac{52}{27} \,\zeta_3^2\BreakII 
\hspace{1cm}
-\frac{3997}{486} \,\zeta_{3}-\frac{5}{6} \,\zeta_{4}+\frac{3875}{243} \,\zeta_{5} \rsb  \,n_f +k^{[2]3}_{2,0}\right\} \left.\rule{0mm}{7mm} \right) 
\nonumber
   \\ 
\hspace{1cm}
= -4 m_s^2 \left(\rule{0mm}{7mm} \right. 1 &+&2.333\, a_s + a_s^2 \left\{
-1.309 \,n_f +23.51\right\}
\nonumber
 \\ &+& a_s^3 \left\{
0.4647 \, n_f^2-32.08 \,n_f +k^{[2]3}_{2,0}\right\} \left.\rule{0mm}{7mm} \right)  {}.  
\label{Pi2m2:exact} 
\end{eqnarray}

Here we have used $a_s = \frac{\alpha_s(Q^2)}{\pi}, m_s = m_s(Q^2) $
and set the normalization scale $\mu^2 = Q^2$; results for generic
values of $\mu$ can be easily recovered with the standard
renormalization group techniques. 
The result for the $ \alpha_s ^4$
terms in $D^{[1]}_0$ has been already presented in
\cite{Baikov:2001aa}, the coefficients of the $ \alpha_s ^4$ and $
\alpha_s ^3$ terms in in $D^{[1]}_2$ and $\Pi^{[2]}_2$ respectively
are new.

\section{Implications for the $ \alpha_s ^4$ predictions and phenomenological analysis}

\subsection{Massless case}

FAC (Fastest Apparent Convergence) and PMS (Principle of Minimal
Sensitivity) methods are both based eventually on the concept  of
scheme-invariant properties and the idea of the choice of an
"optimal" scheme to provide better convergence of the resulting
perturbative series.  For both methods the optimal scheme depends on
the physical observable we are dealing with. With FAC it should be a
scheme which minimizes (set to zero by construction) all the terms of
order $ \alpha_s ^2$ and higher, while the PMS scheme is fixed  by the requirement
that the perturbative expansion for the observable is as insensitive as
possible to a change in the scheme. The assumption that a 
renormalization scheme is in a sense  optimal sets certain constraints on not yet
computed higher order corrections in any other scheme. These
constraints can be used to "predict" (at least roughly) the magnitude of
these corrections.

For the function $D^{[1]}_0$ the result is known since long from 
Ref.~\cite{KS} (see Table~1, column 5).
{}From the three entries corresponding to $n_f = 3,4$ and $5$ one easily
restores the FAC/PMS  prediction   for the $n_f$ dependence
of the $ \alpha_s ^4$ term in  $D^{[1]}_{0}$  (the term of order $n_f^3$ was fixed to its
computed value):
 \begin{equation}                                      
d^{[1]4}_{0} = 127.6 -  44.2 \,    n_f + 3.64 \,  n_f^2   - 0.0100928 \,  n_f^3 
d^{[1]4}_{0} = 127.58 - 44.211 \,  n_f + 3.6439 \,  n_f^2   - 0.0100928 \,  n_f^3 
{}.
\label{K4:nf}
 \end{equation} 
(Note that   FAC and PMS predictions happen to coincide for the $ \alpha_s ^4$ term.)
\begin{table}[ht]
\renewcommand{\arraystretch}{1.3}
\begin{center}
\begin{tabular}{|c|c|c|c|c|c|}
\hline
$n_f$ & $d_{3} ^{\rm exact}$ & $d_3^{\rm FAC}$ & $d_3^{\rm PMS}$ &
$d_4^{\rm FAC/PMS}$  & $d_5^{\rm FAC}$\\
\hline\hline
3 & $ 6.371$ & $ 5.604$ & $ 6.39 $ & $ 27 \pm 16 $ & $ 145 \pm 100$\\
\hline
4 & $ 2.758$ & $ 4.671$ & $ 5.26 $ & $ 8  \pm 28 $ & $ 40  \pm  160$\\
\hline
5 & $ -0.68$ & $ 3.762$ & $ 4.16 $ & $ -8 \pm 44 $ & $ -3   \pm 230$\\
\hline
\end{tabular}
\end{center}
\caption{\label{t1}
Estimates for  the  coefficients $d_3 = d^{[1]3}_{0}$ and $d_4 = d^{[1]4}_{0}$ 
in the function    $D^{[1]}_0 $
based on  FAC and PMS optimizations. 
The estimate for  of $d^{[1]4}_{0}$ employs the exact value of $d^{[1]3}_{0}$.
The last column contains the FAC predictions for the coefficient  $d_5^{\rm FAC}$
which was obtained assuming the value for $d^{[1]4}_{0}$ as given in 
the fifth column; the corresponding uncertainties  have been  estimated as 
described in the  text.} 
\end{table}

It is  interesting to compare the FAC/PMS predictions  
for the $n_f$ dependence of coefficient $d^{[1]3}_{0}$ with the exact result 
given in  eq.~\ref{D1m0:exact}. The results  of both estimates are 
 \begin{equation}                                        
d^{[1]3}_{0}(FAC) =   8.54 - 1.013 \,  n_f + 0.0116 \,  n_f^2
\label{d1m0as3:FAC}
{},
 \end{equation} 
 \begin{equation}                                    
d^{[1]3}_{0}(PMS) =   9.93 - 1.23 \,  n_f + 0.0125 \,  n_f^2 
\label{d1m0as3:PMS}
{}.
 \end{equation} 

The comparison  of the complete $ \alpha_s ^3$ and partial $ \alpha_s ^4$ results with 
FAC and PMS  estimates leads to the following observations:
\begin{itemize}
\item  Starting from $ \alpha_s ^3$, the leading in $ \alpha_s $ {\em and} $n_f$ terms of
order $ \alpha_s ^3 n^2_f$ and $ \alpha_s ^4 n_f^3$ 
 are numerically quite small (at least for $n_f \le 6$)
and, thus, should have a negligibly small influence on the
coefficients of the $ \alpha_s $ expansion.  On the other hand, the term 
subleading in $n_f$, say,  of order $ \alpha_s ^3 n_f$ is comparable in size
with the term of order
$ \alpha_s ^3 n_f^0$. 
Similarly, the $ \alpha_s ^4 n_f^3$ term is significantly smaller than  the 
$ \alpha_s ^4 n_f^2$ one,  whereas the $ \alpha_s ^4 n_f$ and $ \alpha_s ^4 n_f^0$
terms are expected to be of similar magnitude.   

\item  In general FAC/PMS methods correctly reproduce sign and order of
magnitude of the higher order coefficients. The agreement is getting
better when the coefficients happen to be large, as is the case for
the quadratic mass corrections (see below).

\item  Taken separately, the FAC/PMS estimates of the coefficients of
the $n_f$ expansion could deviate rather strongly from the true
result.  However, for a given $n_f$, the deviation in the predicted
value of the full $O( \alpha_s ^n)$ term tends to be significantly smaller
than what could be expected from summing individual terms of the $n_f$
expansion.  In addition, for the particular point $n_f = 3$  very good
agreement is observed.

To illustrate this feature, let us consider the worst case: the
FAC/PMS prediction for the $ \alpha_s ^3 n_f$ term in the function
$d^{[1]}_0$. Here the ratio of the exact result relative to  the predicted one
is quite large (about 4).  Without  knowledge of the $ \alpha_s ^3 n_f^0$
contribution one would expect that the uncertainty of the prediction
for the full $ \alpha_s ^3$ coefficient should be at least  around
\[
(d^{[1]3}_{0,1}|_{exact} -d^{[1]3}_{0,1}|_{FAC}) \ n_f .\]
 For $n_f =
3,4,5 $ this  amounts to 9, 12 and 15, which should be compared to the
corresponding differences of the full order $O( \alpha_s ^3)$ (summed over
all contributing power of $n_f$) coefficients, viz. $1, 2, 4$. This
example demonstrates that the deviation of FAC/PMS predictions for
subleading in $n_f$ terms may well serve as conservative
estimate  of the accuracy in  the prediction of the
complete  terms of  orders $ \alpha_s ^3$ and $ \alpha_s ^4$.
\end{itemize}

These observations   motivate the     
assumption  that the prediction for the  coefficient
$d^{[1]4}_0 
(\equiv d^{[1]4}_{0,0} + d^{[1]4}_{0,1}n_f +d^{[1]4}_{0,2}n_f^2
+  d^{[1]4}_{0,3}n_f^3)
$
should also be correct within 
\[ 
\pm (d^{[1]4}_{0}|_{exact} -d^{[1]4}_{0}|_{FAC}) \ n_f^2 
\]
Thus, in  our  phenomenological  analysis of the $\tau$-lepton 
decays  the  estimate   
 \begin{equation} 
d^{[1]4}_0 |_{n_f=3}= 27 \pm 16 
\label{d1m0as4nf3:num}
 \end{equation} 
will be used.  On the basis of this improved estimate and the four loop $\beta$ 
function \cite{vanRitbergen:1997va} one may even speculate about the $\alpha_s^5$ term 
(whose exact evaluation is completely out of reach  in the foreseeable  future).
Following the discussion of  Kataev and Starshenko \cite{Kataev:1994wd}, one obtains
 \begin{equation} 
d_0^{[1]5}|_{n_f=3} = 145 \pm 100 
\label{d1m0as5nf3:num}
{},
 \end{equation} 
not far from the previous estimates of Ref.~\cite{Kataev:1994wd}.   The variation of  
$d^{[1]4}_0$ by $\pm 16$ leads to the  variation of $ d^{[1]5}_0 $  by $\pm 100$.
For other values of $n_f$, the corresponding predictions can be  
obtained in the same way. They  are listed in  Table~1.

The FAC/PMS prediction (\ref{K4:nf}) for the 
$n_f$ dependence of the coefficient
$d^{[1]4}_{0}$ does not take into account  the available knowledge  of
the corresponding $n_f^2$ part. 
One can easily include this by fitting 
the FAC/PMS predictions for only ${\it two}$ values of $n_f$ with 
a ${\it linear}$  function of $n_f$. As  a result one 
obtains   \footnote{We thank Matthias  Steihnauser 
and Robert  Harlander for a useful  discusson of this
point.}
(we have boxed the predicted coefficients in order to separate them 
clearly from the input)
\bea
d^{[1]4}_{0} (  \mathrm{FAC/PMS} , n_f = 3,4 ) &=&
\fbox{$105.7 - 31.8\,  n_f$}  + 1.875 \, n_f^2   -0.01009 \, n_f^3 
{},
\label{3,4->5}
\\
d^{[1]4}_{0} (  \mathrm{FAC/PMS} , n_f = 4,5 ) &=&
\fbox{$107.7 - 32.3\,  n_f$}  + 1.875 \, n_f^2   -0.01009 \, n_f^3 
{},
\label{4,5->3}
\\
d^{[1]4}_{0} (  \mathrm{FAC/PMS} , n_f = 3,5 ) &=&
\fbox{$106.4 - 32.0 \,n_f$}  + 1.875 \, n_f^2   -0.01009 \, n_f^3 
{}.
\label{3,5->4}
\eea
One could now perform a self-consistency check of Table I by 
predicting, say,   $d^{[1]4}_{0}$ for $n_f=5$ from eq.~(\ref{3,4->5}). 
The result --  (-7.5) --  is compared   successfully to the 
value listed in the table, viz.  -8. 
The corresponding predictions from eqs.~(\ref{4,5->3}) and (\ref{3,5->4})
are also  in very good  agreement to Table~I.

An instructive example of how knowledge and inclusion of the
subleading $n_f$ term can improve FAC/PMS predictions is provided by
the (exactly known) coefficient $d^{[1]3}_{0}$. Indeed, assuming the
knowledge of
$d^{[1]3}_{0,2}$ and $d^{[1]3}_{0,1}$  and using the values of 
$d_3^{\mathrm{FAC}}$  and $d_3^{\mathrm{PMS}}$ 
(at $n_f = 3$) as given by Table 1,  one  easily arrives at 
 \begin{equation}                                        
d^{[1]3}_{0}(\mathrm{FAC}) =   \fbox{17.48} -4.216\,  n_f + 0.08621\, n_f^2
\label{d1m0as3:FAC:nf->3}
{}
 \end{equation} 
and
 \begin{equation}                                    
d^{[1]3}_{0}(\mathrm{PMS}) =   \fbox{18.27} -4.216\,  n_f + 0.08621\, n_f^2
\label{d1m0as3:PMS:nf->3}
{},
 \end{equation} 
which should be compared to the exact value $d^{[1]3}_{0,0} = 18.24$.
Repeating the same  exercise for $n_f=4,5$ 
we get 20.16 (FAC) and  20.75 (PMS) for $n_f=4$ as well as  $22.69$ (FAC) 
and 23.08 (PMS) for $n_f=5$  respectively.

Bearing in mind that in many cases FAC/PMS predictions made for $n_f= 3$ 
are in  better agreement with the exact results (see, e.g. Table~I and tables
from Ref.~ \cite{Kataev:1994wd}), we  suggest  eq.~(\ref{3,4->5}) as the best FAC/PMS
prediction for the constant term  and the term  linear in $n_f$  in the coefficient
$d^{[1]4}_{0}$.

Eqs. (\ref{d1m0as4nf3:num},\ref{d1m0as5nf3:num}) can be used to predict 
$R_{\tau}$  in the massless limit  first  in 
fixed order  perturbation theory (FOPT) 
 \begin{eqnarray} 
&R_{\tau}^{\mathrm{FOPT}} =  &
\label{Rtau:FO}
\\
& 3 \left(  
1 + \, a_s+ 5.202 \, a^2_s+ 26.37 \, a_s^3 +  a_s^4 \ (105 \pm 16) +
 \, a_s^5 (138 \pm 230 \pm 100) 
    \right)
&
   \nonumber  
{}.
 \end{eqnarray} 
Here  $a_s= \alpha_s(M_{\tau})/\pi$. 
The first uncertainty in the $ \alpha_s^5$  term comes from that of 
$d^{[1]4}_0$  while the second is our estimation of the error in 
the very  coefficient  $d^{[1]5}_0$. (Of course, within this approach 
they are strongly correlated).  

Similarly, we can use  "contour improved" (CI)  formulae \cite{tau:resum,DP} 
(assuming as  reference value $ \alpha_s (M_{\tau}) = 0.334$ \cite{exp1})
to get 
 \begin{equation} 
R_{\tau}^{CI} = 3 \left(  
1 + 1.364 \, a_s+ 2.54 \, a_s^2 + 9.71 \, a_s^3 
 + 1.31 \, a_s^4 d^{[1]4}_0  + 0.95 \, a_s^5  d^{[1]5}_0
    \right) 
{}
\label{Rtau:CIa}
 \end{equation} 
or, equivalently,  
 \begin{equation} 
R_{\tau}^{CI} = 3 \left(  
1+ 1.364\, a_s+ 2.54 \, a_s^2 + 9.71 \, a_s^3 + \, a_s^4 (35 \pm 20) + a_s^5 (138  \pm 95)
                    \right) 
\label{Rtau:CI}
{}.
 \end{equation} 

Let us compare our new value for the coefficient $d^{[1]4}_0$
in (\ref{d1m0as4nf3:num}) with the ones used in extracting 
$\alpha_s(M_{\tau})$ from $\tau$ data by the OPAL \cite{exp2} and ALEPH
\cite{exp1} collaborations, namely 
 \begin{equation}  d^{[1]4}_0 |_{n_f=3} = 25 \pm 25
\ \mbox{(OPAL)} , \ \ \ 50 \pm 50 \ \mbox{(ALEPH)} 
\label{d4:OPAL+ALEPH} {}.
  \end{equation} 
While OPAL's central value is basically the same as ours, their error
bar is somewhat larger. In the case of ALEPH both the
central value and the uncertainty assigned to it are significantly
larger than our numbers.  In this connection, we would like to stress
that eq.~(\ref{d1m0as4nf3:num}) utilizes completely new non-trivial
information  given in (\ref{D1m0:exact}): the subleading term  
in $n_f$ term of order  $ \alpha_s ^4$.  

It is of interest to see in detail which accuracy in the determination
of $\alpha_s(M_{\tau})$ one could achieve assuming
eqs. (\ref{d1m0as4nf3:num},\ref{d1m0as5nf3:num}).
Let us 	 introduce the 
quality $\delta_P$ as follows:  
 \begin{equation} 
R_{\tau S=0} = \
\frac{\Gamma(\tau \to h_{S=0} \nu)}{\Gamma(\tau \to l \bar{\nu}\nu)}
=  |V_{ud}|^2 S_{EW} R_{\tau}
{},
\label{delta_exp}
 \end{equation} 
with 
\[
R_{\tau} = 3 (1 + \delta_P + \delta_{EW} + \delta_{NP}) 
{}.
\]
The first term here is the parton result,  the second  stands for pQCD effects.
The non-perturbative corrections  represented by  $\delta_{NP}$
happens to  be rather small $\delta_{NP} = -.003 \pm 0.003$ 
(see, e.g. \cite{NarPic88}).
Here the flavour mixing matrix element  
$|V_{ud}|^2 = 0.9475 \pm 0.0016$ \cite{PDG}.
The factor $S_{EW} = 1.0194 $  is the electroweak correction   
which collects the large logarithmic terms 
\cite{ewcorr1}, 
while $\delta_{EW} = 0.001$ is an additive electroweak correction \cite{ewcorr2}.  
Using for definiteness the result of ALEPH~\cite{exp1}
 \begin{equation} 
R_{\tau S=0} = 3.492 \pm 0.016 
{},
\label{Rtau:ALEPH}
 \end{equation} 
one arrives at
 \begin{equation} 
\delta_P^\mathrm{exp} =  0.207 \pm 0.007
\label{delta_P:exp}
{}.
 \end{equation} 
To get a value for $\alpha_s(M_{\tau})$ one should simply
fit $\delta_P^\mathrm{exp}$ against $R_{\tau}/3 - 1$ as given by eq.~(\ref{Rtau:FO}) or
by eq.~(\ref{Rtau:CI}) to get a result corresponding 
FOPT or  CIPT (Fixed Order or Contour Improved PT). 

Unfortunately, there is no 
unique way to assign a
theoretical uncertainty $\delta { \alpha_s }$ to the obtained value of  
$\alpha_s(M_{\tau})$. In the literature one finds several suggestions.
Let us consider them in turn. 

\begin{enumerate}

\item  $\delta { \alpha_s }$ is  a half of the shift in $\alpha_s$ induced by the last
       fully  computed
       term in the PT (that is by the one of order $ \alpha_s ^3$ at present).

\item    $\delta { \alpha_s }$ is equal to the  change  in $\alpha_s$ caused by varying the normalization point
     $\mu$ around $M_{\tau}$, typically within the range of 1.1--2.5 GeV.

The corresponding results read (terms of order $ \alpha_s ^4$ and higher in 
 eq.~(\ref{Rtau:FO})  and in    the $D$-function have been set to zero)
\begin{eqnarray} 
\alpha_s ^{\mathrm{FOPT}}(M_{\tau}) &=&   0.345 \pm    (0.025| 0.037  ) 
{},
\label{asFOas3}
\\
 \alpha_s ^{\mathrm{CIPT}}(M_{\tau}) &=&    0.364  \pm  (0.012| 0.021   ) 
{}.
\label{asCIas3}
 \end{eqnarray} 
Here the first(second) value in brackets  correspond to the
use of the first(second) suggestion for the error  estimation.   
After evolution from $M_{\tau}$  to $M_Z$ this corresponds to
 \begin{eqnarray} 
 \alpha_s ^{\mathrm{FOPT}}(M_{Z}) &=&   0.1209  \pm   (0.0024 | 0.0037 ) 
{},
\label{asFOas3_MZ}
\\
 \alpha_s ^{\mathrm{CIPT}}(M_{Z}) &=&   0.1229   \pm  (0.0011 | 0.0020 ) 
{}.
\label{asCIas3_MZ}
 \end{eqnarray} 
\item  
 $\delta { \alpha_s }$ is  equal to the  change in $\alpha_s$ caused by the uncertainty in the 
    predicted (that is not yet completely known) higher order 
    terms in the perturbative series for $R_{\tau}$;

\item  $\delta { \alpha_s }$ is  a  half of the difference  in the    $\alpha_s(M_{\tau})$  as 
    obtained within FOPT and CIPT. This difference 
    comes from different handling of higher order terms. 
\end{enumerate}
In order to quantify the error estimates according to 3. and 4.  we
have shown in Table~\ref{t2} the results for $ \alpha_s (M_{\tau})$ obtained with
various choices for $d^{[1]4}_0$, $d^{[1]5}_0$ and $\mu$. The entries
with the choices $\pm 100$ for the coefficients illustrate the large
change  in $ \alpha_s $ which would result from a failure of PMS and
FAC once higher order terms are included. 
For the plausible values of $d_0^{[1]4}$ and $d_0^{[1]5}$ we
observe an significant  decrease of the $\mu$ dependence  after
inclusion  of  additional terms in the $ \alpha_s $ series.

Our final  predictions
for $ \alpha_s ^{\mathrm{FOPT}}(M_{\tau})$ and $ \alpha_s ^{\mathrm{CIPT}}(M_{\tau})$
are given in the first column of Table~\ref{alpha_fin}, 
together with experimental error\footnote{
According to eq.~(\ref{delta_P:exp}) the latter
includes  (small) uncertainties in the values of $|V_{ud}|$ and the nonperturbative
correction $\delta_{NP}$  in addition to the experimental error per se as displayed
in~eq.~(\ref{Rtau:ALEPH}).}
and  the combined  theory uncertainty.  The values of 
theory uncertainties are listed separately  in columns 3,4 and 5.
The corresponding values at the  scale
of $M_Z$ are
 \begin{eqnarray} 
 \alpha_s ^{\mathrm{FOPT}}(M_{Z}) &=&   0.1192  \pm   0.0007 \pm   0.002   
{},
\label{asFOas3_MZ_fin}
\\
 \alpha_s ^{\mathrm{CIPT}}(M_{Z}) &=&   0.1219   \pm 0.001   \pm 0.0006  
{}.
\label{asCIas3_MZ_fin}
 \end{eqnarray} 
\begin{table}[ht]
\renewcommand{\arraystretch}{1.3}
\begin{center}
\begin{tabular}{|c|c|c|c|} 
\hline
$d^{[1]4}_0$  & $ \alpha_s ^4$ & $d^{[1]5}_0$ & $ \alpha_s ^5$ 
\\ 
\hline
 27 & $0.331 \pm 0.02$ & 145 & $0.33 \pm 0.02$ 
\\ 
\cline{2-2}\cline{4-4} 
  &   $0.357 \pm 0.009$ &  &$0.354 \pm 0.004$ 
\\ 
\hline
 43 & $0.329 \pm 0.02$ & 245 & $0.324 \pm 0.01$ 
\\ 
\cline{2-2}\cline{4-4} 
  &   $0.353 \pm 0.01$ &  &$0.348 \pm 0.005$ 
\\ 
\hline
 11 & $0.333 \pm 0.02$ & 45 & $0.335 \pm ?$ 
\\ 
\cline{2-2}\cline{4-4} 
  &   $0.361 \pm 0.008$ &  &$0.360 \pm 0.002$ 
\\ 
\hline
 100 & $0.324 \pm 0.02$ & 100 & $0.314 \pm 0.01$ 
\\ 
\cline{2-2}\cline{4-4} 
  &   $0.340 \pm 0.01$ &  &$0.338 \pm 0.007$ 
\\ 
\hline
 -100 & $0.348 \pm ?$ & -100 & $0.349 \pm ?$ 
\\ 
\cline{2-2}\cline{4-4} 
  &   $0.398 \pm 0.03$ &  &$0.401  \pm ?    $ 
\\
\hline 
\end{tabular}
\end{center}
\caption{\label{t2}
The predicted value of $ \alpha_s (M_{\tau})$ in dependence of 
chosen  values for the coefficients 
$d^{[1]4}_0$, $d^{[1]5}_0$.
The second and the forth  columns differ in the the number of terms in the
perturbative series included. 
The upper value of $ \alpha_s $ is the one predicted within FOPT, the
lower  corresponds to CIPT. The uncertainty in the value of $ \alpha_s $ corresponds 
to changing the  normalization point $\mu$ as follows $\mu^2/M_{\tau}^2 = 0.4 - 2.$
The  entry with question mark means that an  equation for $ \alpha_s (\mu)$ does
not have a solution for some  value of $\mu$ within the interval. 
} 
\end{table}

Thus we observe  that the total uncertainty based on a combination of 
not  yet calculated  higher order terms, $\mu$-dependence  and scheme 
dependence  is reduced, once $ \alpha_s ^4$  terms are available. However,  the 
difference  between FOPT and  CIPT results   of roughly 0.02 is a  remaining,
at the moment irreducible uncertainty\footnote{
This is in agreement to the analysis of
Ref.~\cite{Groote:1997cn}, where it  has been concluded  that 
``the resummed values of $\alpha_s$ from $\tau$
decay lie outside the convergence radii and can therefore not be obtained
from a power series expansion. Regular perturbation series do not converge
to their resummed counterparts. The experimental value of $R_{\tau}$ appears
to be too large for a fixed order perturbation analysis to apply''.
See also \cite{krajals}.}.

\begin{table}[ht]
\renewcommand{\arraystretch}{1.3}
\begin{center}
\begin{tabular}{|c|c|c|c|c|c|} 
 \hline 
Method & $ \alpha_s (M_{\tau})$ & $\Delta \, \delta^\mathrm{exp}_P$ & 
 $\Delta \, \mu $ & $\Delta \, d_0^{[1]4}$ & $\Delta \, d_0^{[1]5}$
 \\ 
 \hline 
FOPT & $0.330 \pm 0.006 \pm 0.02$ & 0.006 & 0.019 & 0.0045 & 0.0011
 \\ 
 \hline 
CIPT & $0.354\pm 0.009 \pm 0.006  $ & 0.009 & 0.0036 & 0.0042 & 0.0019
\\
\hline 
\end{tabular}
\end{center}
\caption{\label{alpha_fin}
The  value of $ \alpha_s (M_{\tau})$ obtained with 
$ \delta^\mathrm{exp}_P$,  $d_0^{[1]4}$ and $d_0^{[1]5}$ fixed to their 
central values according to  
eqs.~(\ref{delta_P:exp},\ref{d1m0as4nf3:num},\ref{d1m0as5nf3:num})
together with corresponding errors.
} 
\end{table}

It is thus of interest to study this difference as a function of
$m_{\tau}$. In practice, this could be applied to sum rules for
spectral functions as determined in $e^+ e^-$ annihilation.

Therefore, let us consider a hypothetical case of $\tau$ lepton with
the mass equal 3 GeV.  Assuming $ \alpha_s (1.77 \,\mathrm{GeV}) = 0.334$ and running
this value to 3 GeV via standard 4-loop evolution equation one gets
$ \alpha_s (3 \,\mathrm{GeV}) = 0.2558$ and predicts 
\[ \delta_P^\mathrm{exp} = 0.1353 \]
which
corresponds to  eq.~(\ref{Rtau:FO})  with the $ \alpha_s ^4$ and $ \alpha_s ^5$ terms fixed
to their central values (see eqs.~(\ref{d1m0as4nf3:num}--\ref{d1m0as5nf3:num})).  \
Let us now  investigate the results for $ \alpha_s $ 
and  the theory error  that would  result from $\delta_P^\mathrm{exp} = 0.1353$
as a starting point.
The corresponding analogs
of eqs. 
(\ref{asFOas3}--\ref{asCIas3_MZ})
and Table~(\ref{t2}) are displayed
below as eqs. ((\ref{asFOas3_3GeV},\ref{asCIas3_3GeV})) and Table
(\ref{t2_Mtau_3}) correspondingly.  The difference between FOPT and CIPT
decreases significantly, and this remains true even after extrapolating
to $ \alpha_s (M_Z)$.  
 \begin{eqnarray} 
 \alpha_s ^{\mathrm{FOPT}}(3 \mbox{GeV}) &=&   0.263  \pm   (0.013|0.014   ) 
{},
\label{asFOas3_3GeV}
\\
 \alpha_s ^{\mathrm{CIPT}}(3 \mbox{GeV}  &=&    0.265  \pm  (0.005| 0.008  ) 
{},
\label{asCIas3_3GeV}
\\
 \alpha_s ^{\mathrm{FOPT}}(M_{Z}) &=&   0.1198  \pm   (0.002|0.003   ) 
{},
\label{asFOas3_3GeV_MZ}
\\
 \alpha_s ^{\mathrm{CIPT}}(M_{Z}) &=&    0.1203  \pm  (0.009| 0.0016  ) 
{}.
\label{asCIas3_3GeV_MZ}
 \end{eqnarray} 

\begin{table}[ht]
\renewcommand{\arraystretch}{1.3}
\begin{center}
\begin{tabular}{|c|c|c|c|} 
\hline
$d^{[1]4}_0$  & $ \alpha_s ^4$ & $d^{[1]5}_0$ & $ \alpha_s ^5$ 
\\ 
\hline
 27 & $0.256 \pm 0.007$ & 145 & $0.256 \pm 0.003$ 
\\ 
\cline{2-2}\cline{4-4} 
  &   $0.262 \pm 0.004$ &  &$0.261 \pm 0.002$ 
\\ 
\hline
 43 & $0.256 \pm 0.007$ & 245 & $0.254 \pm 0.003$ 
\\ 
\cline{2-2}\cline{4-4} 
  &   $0.26 \pm 0.004$ &  &$0.259 \pm 0.002$ 
\\ 
\hline
 11 & $0.257 \pm 0.006$ & 45 & $0.258 \pm 0.005$ 
\\ 
\cline{2-2}\cline{4-4} 
  &   $0.264 \pm 0.003$ &  &$0.263 \pm 0.001$ 
\\ 
\hline
 100 & $0.253 \pm 0.008$ & 100 & $0.249 \pm 0.004$ 
\\ 
\cline{2-2}\cline{4-4} 
  &   $0.255 \pm 0.008$ &  &$0.254 \pm 0.002$ 

\\ 
\hline
 -100 & $0.264 \pm 0.01$ & -100 & $0.277 \pm \,  ?$ 
\\ 
\cline{2-2}\cline{4-4} 
  &   $0.279 \pm 0.02$ &  &$0.28 \pm \, ?$ 
\\
\hline 
\end{tabular}
\end{center}
\caption{\label{t2_Mtau_3}
The predicted value of $ \alpha_s (M_{\tau})$ in dependence of 
chosen  values for the coefficients 
$d^{[1]4}_0$, $d^{[1]5}_0$ for a hypothetical case
of $M_{\tau} = 3$ GeV.  
The third and the forth  columns differ in the the number of terms in the
perturbative series included. 
The upper value of $ \alpha_s $ is the one predicted within FOPT, the
lower  corresponds CIPT. The uncertainty in the value of $ \alpha_s $ corresponds 
to changing the  normalization point $\mu$ as follows 
$\mu^2/M_{\tau}^2 = 0.4 - 2.$
The entry with question mark means that an  equation for $ \alpha_s (M_{\mu})$ does
not have a solution for some  value of $\mu$ within the interval. 
} 
\end{table}

The same coefficients 
$d^{[1]4}_0 $ and $ d^{[1]5}_0$  
can be used  to predict \cite{KS,Kataev:1994wd}
(the non-singlet part of) corrections of orders $\alpha_s^4$ and 
$\alpha_s^5$  to the $R$-ratio in Z  decays:

\[
R(n_f=5) = 1+ a_s+ 1.409 \, a_s^2 -12.77 \, a_s^3 + (-97 \pm 44) \, a_s^4 +(76 \pm 230) \, a_s^5
{}.
\]

It is also of interest to display $R(s)$ for $n_f =3$ and 4,
which is accessible at $e^+ e^-$ colliders at lower energies.

\[
R(n_f = 4) = 1+ a_s+ 1.525 \, a_s^2 - 11.52 \, a_s^3 + (-112 \pm 30) \, a_s^4 
+(-245 \pm 160) a_s^5
{}.
\]

\[
R(n_f = 3) = 1+ \, a_s+ 1.640 \, a_s^2 -10.28 \, a_s^3 + (-129 \pm 16) \, a_s^4 + (-635 \pm 100) a_s^5
{}.
\]
Our results are close to those of \cite{KS,Kataev:1994wd},
they employ, however, the additional information from
\cite{Baikov:2001aa,vanRitbergen:1997va}.

These formulae  demonstrate rather
good convergency for $n_f =5 $ and a reasonably good one for $n_f =3$
and $4$ if our predictions for the coefficients $d^{[1]4}_0$,
$d^{[1]5}_0$ deviate from the true values within the assumed 
error margins.

\subsection{Quadratic mass corrections}
Let us first discuss  the function $ D^{[1]}_2 $. 
The FAC/PMS predictions can be easily obtained  following   
\cite{CheKniSir97}; they   are listed  in  Table~\ref{t3}. 

\begin{table}[ht]
\renewcommand{\arraystretch}{1.3}
\begin{center}
\begin{tabular}{|c|c|c|c|c|}
\hline
$n_f$ & $d_{3} ^{\rm exact}$ & $d_3^{\rm FAC}$ & $d_3^{\rm PMS}$ &
$d_4^{\rm FAC/PMS}$ \\
\hline\hline
3 & $ 107.5$ & $ 89.86$ & $ 91.17 $ & $ 1200 \pm 400$ \\
\hline
4 & $ 94.37$ & $ 79.13$ & $ 80.11 $ & $ 950  \pm 300$  \\
\hline
5 & $ 81.54$ & $ 68.66$ & $ 69.33 $ & $ 750  \pm  200$  \\
\hline
\end{tabular}
\end{center}
\caption{\label{t3}
Estimations of the  coefficients $d_3 = d^{[1]3}_{2}$ and $d_4 = d^{[1]4}_{2}$ 
in the function    $D^{[1]}_2$ based on the FAC and PMS optimizations.
The estimation of $d^{[1]4}_{2}$ employs the exact value of $d^{[1]3}_{2}$.}
\end{table}

We again restore the $n_f$ dependence of the coefficient
$d^{[1]4}_{2}$ as predicted by FAC/PMS:
 \begin{equation} 
d^{[1]4}_{2}(\mbox{FAC/PMS})   = 1931.44 - 281.956 \,  n_f + 9.0294 \,  n_f^2 - 0.0156289 \,  n_f^3
\label{d1m2as4:FAC,PMS}
{}
 \end{equation} 
as well as  that of $d^{[1]3}_{2}$:
 \begin{equation} 
d^{[1]3}_{2}(\,\mbox{FAC}) =  \ \ 123.654 - 11.6638 \,  n_f + 0.133293 \,  n_f^2
\label{Pi2m2as3:FAC}
{},
 \end{equation} 
 \begin{equation}                                  
d^{[1]3}_{2}(\,\mbox{PMS})           = \ \ 125.975 - 12.0028 \,  n_f + 0.134769 \,  n_f^2
\label{Pi2m2as3:PMS}
{}
 \end{equation} 
to be compared with
 \begin{equation} 
d^{[1]3}_{2}(\,\mbox{exact}) = 148.978 - 14.3097 \,  n_f     +0.16463 \,  n_f^2
\nonumber
{}.
 \end{equation} 
The  comparison of estimates and  exact results reveals a picture
qualitatively similar to the massless case but with some modifications.  
A few   important observations are in order.

1.  All three terms of the $n_f$ expansion of $d_2^{[1]3}$ are
successfully predicted within about 20 \% accuracy.

2.  Unlike the massless case the agreement between FAC/PMS predictions
for the coefficient $d_2^{[1]3}$ for $n_f = 3,4,5$ and the
corresponding exact numbers is within the range of 15\%--20\%. On the
other hand, the estimation of accuracy of the  $ \alpha_s ^3$ fixed $n_f$
predictions obtained {\em exclusively} from the knowledge of
the subleading  contribution of $O(  \alpha_s ^3 n_f )$  is of the right
order of magnitude but somewhat less.  All this is probably a
consequence of a significantly larger $n_f$ independent contribution.

3. At ${\mathcal{O}}( \alpha_s ^4)$  the exact result for the full
coefficient 
\[
d_2^{[1]4} =d^{[1]4}_{2,0} + d^{[1]4}_{2,1} n_f 
+ d^{[1]4}_{2,2} n_f^2 + d^{[1]4}_{2,3} n_f^3
\]
is unknown, apart from its leading and subleading terms in $n_f$. The
prediction  for the subleading coefficient 
$d^{[1]4}_{2,2}  = 9.0 $ is by 50\% larger  than the exact 
value  6.07. 
The predicted values
for $d^{[1]4}_{2,0}$ and $d^{[1]4}_{2,1}$ are {\em very} large. In
view of this largeness, the estimate $ (d^{[1]4}_{2,2}|_{exact}
-d^{[1]4}_{2,2}|_{FAC}) \ n_f^2 $ (= 30, 50, 80 for $n_f = 3,4,5$
respectively) looks somewhat too optimistic. Therefore  we 
assign a conservative 30\% uncertainty to the fixed $n_f$
predictions listed in the fifth column of the Table~\ref{t3}.

\begin{table}[ht]
\renewcommand{\arraystretch}{1.3}
\begin{center}
\begin{tabular}{|c|c|c|c|}
\hline
$n_f$ &  $k_3^{\rm FAC}$ & $k_3^{\rm PMS}$ &
$k_4^{\rm PMS}$ \\
3 &   $ 199.1$ & $ 200 \pm 60  $ & $ 2200 \pm 1500 $\\
\hline
4 &   $ 171.2$ & $ 170 \pm 50 $  & $ 1800  \pm  1100 $\\
\hline
5 &    $ 144.7$ & $ 145 \pm 40 $ & $ 1400 \pm  900 $\\
\hline
\end{tabular}
\end{center}
\caption{\label{t5}
Estimates of the  coefficients $k_3 = k^{[2]3}_{2}$ and $k_4 = k^{[2]4}_{2}$ 
in the function    $\Pi^{[2]}_2$
based on the FAC and PMS optimizations.
The estimate of $k^{[1]4}_{2}$ employs the predicted value for $k_3$;
the corresponding uncertainties  include only the ones induced by $k_3$.} 
\end{table}

At last, we repeat  the analysis for the function $ \Pi^{[2]}_2$. 
The results of FAC/PMS optimization methods are given in  Table~\ref{t5}. 
Using the values of $k_3^{\rm PMS}$ from the  table for
$n_f=3,4,5$ we reconstruct the corresponding full $n_f$ dependence:
\[ k^{[1]3}_{2} = 294.472 - 33.2429 \,  n_f + 0.696598 \,  n_f^2
{}.  
\]
The  comparison with the known terms of order $n_f$ and $n_f^2$
($- 32.0843 \,  n_f , \ \ 0.464663 \, nf^2 $) demonstrates a remarkably good
agreement for the subleading  $n_f$ contribution. 
The 50\% error in the predicted value of the $n_f^2$  contribution 
looks  natural as the corresponding coefficient is
small. Following the same line of reasoning as above we  have assigned a 
30\% uncertainty to the $O( \alpha_s ^3)$ fixed $n_f$  result.

To get a general idea about the  size of the $ \alpha_s ^4$ contribution to
eq.~(\ref{Pi2m2:exact}) we used FAC/PMS and the predicted $ \alpha_s ^3$ coefficient. 
The results are listed  in the forth column of Table~\ref{t5}.
 
Let us now consider the effect of $ \alpha_s ^3$ and $ \alpha_s ^4$ corrections on
the determination of the strange quark mass\footnote{For a recent
review of various attempts to extract the strange quark mass from
$\tau$ data see Ref.~\cite{Maltman:2002wb}.}.
The mass correction to the $R_{\tau}$ depend on both functions
$D^{[1]}_2$ and $ \Pi^{[2]}_2$. Let us  use the central ALEPH value 
of $ \alpha_s (M_{\tau}) = 0.334 $
when estimating  the size of the perturbative corrections. For  fixed order
one finds the mass correction to the total rate:
 \begin{eqnarray} 
{\delta}^{00}_{us,2} &=& 
-8\frac{m_s^2}{M_\tau^2}(
1. + 5.33  \, a_s+ 46.0  \, a_s^2  + 284  \, a_s^3  
\nonumber\\ 
&+& 0.75 \, a_s^3   k^{[2]3}_2   
+ a_s^4(723. + 0.25  \ d^{[1]4}_2  + 9.84 \  k^{[2]3}_2  + 0.75\   k^{[2]4}_2 )
  ) 
\nonumber 
\\ 
&{=}&
-8\frac{m_s^2}{M_\tau^2}(1. + 0.567  + 0.520   +   0.521\pm 0.05 + 0.593   )
 \nonumber \\ 
&{=}&
-8\frac{m_s^2}{M_\tau^2}
(3.2  \pm   0.6)
\label{num:disc:1.3}
{},
 \end{eqnarray} 
where in the last line we have assumed the (maximal!) value of 
the ${\cal{O}}(\alpha_s^4)$
term as an estimate of the theoretical  uncertainty
(this convention will  be used also below).

For the  ``contour improved'' series one obtains 
 \begin{eqnarray} 
\tilde{\delta}^{00}_{us,2} &=& 
-8\frac{m_s^2}{M_\tau^2}(
1.44 + 3.65  \, a_s+ 30.9 \, a_s^2  + 72.2  \, a_s^3  + 1.18  \, a_s^3 k_2^{[2],3} 
\nonumber
\\
&{}& \phantom{MMMMMMMxxl}  + a_s^4 ( 0.678 \  d^{[1]4}_2  + 1.06 \  k^{[2]4}_2 )
 ) 
 \nonumber \\  
&{=}&
-8\frac{m_s^2}{M_\tau^2}(1.44 + 0.389 + 0.349   + 0.371 \pm 0.09 + 0.403)
 \nonumber \\ 
&{=}&
-8\frac{m_s^2}{M_\tau^2}
( 2.95 \pm  0.4)
\label{num:disc:2.3}
{}.
 \end{eqnarray}

Now we consider the contributions of spin 1 and spin 0 separately.
The lowest moments ($L = 0$) of the spin-dependent functions depend on
a nonperturbative quantity and, thus, can not be treated
perturbatively in principle \cite{Chetyrkin:1998ej}.

For the spin one part and for $(k,l) = (0,1)$ we find  
 \begin{eqnarray} 
{\delta}^{(1)01}_{us,2} &=& 
-5 \frac{m_s^2}{M_\tau^2}( 
1. + 4.83  a + 35.7  \, a_s^2  + 276. \, a_s^3 +a_s^4 (1350 + d^{[1]4}_2))
                       ) 
 \nonumber \\ 
&{=}&
-5  \frac{m_s^2}{M_\tau^2}
(1. + 0.514  + 0.404   + 0.331 + 0.326)
 \nonumber \\ 
&{=}&
-5 \frac{m_s^2}{M_\tau^2}
(2.58 \pm 0.33)
{}
\label{num:disc:5.3}
 \end{eqnarray} 
and
 \begin{eqnarray} 
\tilde{\delta}^{(1)01}_{us,2} &=& 
-5\frac{m_s^2}{M_\tau^2}(
1.37 + 2.55  a + 16.1  \, a_s^2  + 135 \, a_s^3 + 0.895 \, a_s^4 d_2^{[1]4}) 
 \nonumber \\ 
&{=}&
-5 \frac{m_s^2}{M_\tau^2}(  1.37 + 0.271  + 0.182   + 0.163 + 0.137 )
 \nonumber \\ 
&{=}&
-5  \frac{m_s^2}{M_\tau^2}
(  2.12  \pm  0.14)
{}.
\label{num:disc:.6.3}
 \end{eqnarray} 
Note that spin 1 contribution is determined by the component 
$\Pi^{[1]}$ alone and is known up to third order.
Clearly, this series is decreasing in a reasonable way
(comparable to the behaviour of  $\tilde{\delta}^{00}_{us,2}$)
 and, at the same
time, only moderately  dependent on the improvement 
prescription with
$\tilde{\delta}^{(1)01}_{us,2}/{\delta}^{(1)01}_{us,2} = 0.82$.
On the basis of Eq.~(\ref{num:disc:.6.3})  
this moment might well serve for a reliable $m_s$ determination, with 
a sufficiently  careful interpretation of the theoretical
uncertainty.

The corresponding spin zero part is, per se, proportional to 
$m_s^2$ (not counting non-perturbative,  so-called ``condensate'' contributions)
and thus could be considered as ideal for a 
measurement  of $m_s$. However, the behaviour of the perturbative
series 
 \begin{eqnarray} 
{\delta}^{(0)01}_{us,2} &=& 
\frac{3}{2} \frac{m_s^2}{M_\tau^2}( 
1. + 9.33  a + 110 \, a_s^2  + 1323 \, \, a_s^3 
  \nonumber   \\
&{}&\phantom{MMMMMM}   + a_s^4 (12200 + d_2^{[1]4} + 17.5 k_2^{[2]3})
\nonumber \\ 
&{=}&
\frac{m_s^2}{M_\tau^2}
(1. + 0.992  + 1.24   + 1.59 + 2.16)
 \nonumber \\ 
&{=}&
\frac{3}{2} \frac{m_s^2}{M_\tau^2}
(7.0  \pm 2 )
\label{num:disc:.7.3}
{}
 \end{eqnarray} 
and
 \begin{eqnarray} 
\tilde{\delta}^{(0)01}_{us,2} &=& 
\frac{3}{2}\frac{m_s^2}{M_\tau^2}(
3.19 + 11.2  a + 126.  \, a_s^2  + 289.  \, a_s^3  + 6.63  \, a_s^3  k_2^{[2]3}
  \nonumber  
\\
&{}&\phantom{MMMMMM}   + a_s^4 (2.71 d_2^{[1]4}  +  7.76 k_2^{[2]4})
                         ) 
 \nonumber \\ 
&{=}&
 \frac{3}{2}\frac{m_s^2}{M_\tau^2}( 
3.19 + 1.19  + 1.42  + 1.94  + 2.6 \pm 1.31
                                  )
 \nonumber \\ 
&{=}&
\frac{3}{2}  \frac{m_s^2}{M_\tau^2}( 10.3 \pm 2.6 )
\label{num:disc:.8.3}
{}
 \end{eqnarray} 
shows  a rapid growth of the coefficients. The series is not expected
to provide an accurate prediction for the mass effects.

\section{Summary}

Implications of the newly calculated $\alpha_s^4 n_f^2$ terms together
with the $\alpha_s^4 n_f^3$ terms for an improved extraction of $ \alpha_s $
from the $\tau$ decay are presented.  Arguments are presented in
support of predictions for the remaining terms of order $\alpha_s^4
n_f$ and $\alpha_s^4 n_f^0$ which are based on FAC or PMS
optimization. The complete calculation will lead to a reduction of the
theory uncertainty within the frameworks of FOPT or CIPT down to a
negligible amount.  However, an irreducible difference between the
results from these two schemes of $\delta  \alpha_s (M_{\tau}) \approx 0.02$
corresponding to $\delta  \alpha_s (M_{Z}) \approx 0.002$ persists even
after inclusion of $\mathcal{O}(\alpha_s^4)$ (and even
$\mathcal{O}(\alpha_s^5)$ terms).  Similar investigations, based on data
up to higher energies (e. g. for fictitious heavy lepton of 3 GeV
of for sum rules based on $e^+ e^-$ data) would
lead to significantly smaller errors.

New contributions of orders $\mathcal{O}(m_s^2\alpha_s^4 n_f^2)$ and
$\mathcal{O}(\alpha_s^4 n_f^2)$ to (axial)vector correlators relevant
for the QCD description of the semileptonic $\tau$ decay into hadrons
are obtained.  The moments $R^{(0,0)}$ and $R^{(0,1)}$ are evaluated
separately for spin zero and spin one final states
\cite{Chetyrkin:1998ej}.  The results are tested against predictions
of FAC/PMS optimization methods.  Good agreement is found. This has
motivated us to take the full of FAC/PMS predictions as the basis for
a new extraction of $\alpha_s$ and $m_s$ from $\tau$-decays with
$\mathcal{O}(\alpha_s^4)$ accuracy.  Using $\delta^\mathrm{exp}_P =
0.207$ we find 0.330 and 0.1192 for $ \alpha_s
^{\mathrm{FOPT}}(M_{\tau})$ and $ \alpha_s ^{\mathrm{FOPT}}(M_{Z})$
respectively. In the framework of contour improved evaluation these
values increase to 0.354 and 0.1219 respectively

In contrast to the massless
result, the PT series contributing to the $m_s^2$ dependent part seem
to be barely convergent and the additional higher order terms
seemingly do not lead to any significant improvement of the
theoretical accuracy in the determination of the strange quark mass
from the $\tau $ decays.

A slightly more favorable pattern of convergence is observed for the
moments of the spin one contribution separately.

\section{Acknowledgments}

The authors are grateful to A.A.~Pivovarov for a lot of useful
discussions.  One of us (K.G.~Ch.) acknowledges  useful
correspondence to K.~Maltman and A.L.~Kataev. This work was supported by the
DFG-Forschergruppe {\it ``Quantenfeldtheorie, Computeralgebra und
Monte-Carlo-Simulation''} (contract FOR 264/2-1), by INTAS (grant
00-00313), by RFBR (grant 01-02-16171), by Volkswagen Foundation and
by the European Union under contract HPRN-CT-2000-00149.

\sloppy
\raggedright

\end{document}